\documentstyle[12pt]{article}
\title{Time in quantum gravity and\\
black-hole information paradox}
\author{Hrvoje Nikoli\'c \\
Theoretical Physics Division, Rudjer Bo\v{s}kovi\'{c} Institute, \\
P.O.B. 180, HR-10002 Zagreb, Croatia \\
{\normalsize e-mail: hrvoje@thphys.irb.hr} \\
\makebox[1in]{} \\
}
\date{\today}
\begin{document}
\maketitle
\begin{abstract}
The fact that canonical quantum gravity does not possess a fundamental 
notion of time implies that the theory is unitary in a trivial sense.
At the fundamental level, this trivial unitarity leaves no room for a 
black-hole information loss. Yet, a phenomenological loss of information
may appear when some matter degrees of freedom are reinterpreted as a clock-time.
This explains how both fundamental unitarity and phenomenological 
information loss may peacefully coexist,
which offers a resolution of the black-hole information paradox.
\end{abstract}
\vspace*{0.5cm}
\vspace*{0.9cm} 


\section{Introduction}

The black-hole information paradox \cite{hawk2}
(see, e.g., \cite{gid,har,pres,pag,gid2,str,math,hoss,fabbri} for reviews) is one of the greatest unsolved puzzles in theoretical physics.
One of the proposed solutions is a generalized quantum theory \cite{hartle1,hartle2,nik,nik2}
in which a pure quantum state describes information 
in the whole spacetime, rather than that
on a spacelike hypersurface as in the usual formulation of quantum theory.
Such a generalized state describes correlations between
outgoing Hawking particles remaining after the complete black-hole evaporation
and ingoing Hawking particles existing before that, thus avoiding the information loss.

Such a generalized quantum theory itself is motivated by a requirement 
of explicit spacetime covariance (see also \cite{nikinqi,nikijmpa}). Yet, in this paper we show that 
explicit spacetime covariance and generalized quantum theory are not essential at all. 
We explain how a similar idea works 
in a more conventional canonical quantum gravity. The essential ingredient common to both
the approach in \cite{nik,nik2} and that in the present paper 
is the idea that there is no fundamental notion of time evolution. In both approaches
time is treated as a local quantum observable, and not as a global external parameter.
We argue that such a treatment of time is sufficient to avoid information loss through
the correlations between ``future'' and ``past''.

The paper is organized as follows. Sec.~\ref{SEC2} is a brief review of
some well-established aspects of the problem of time in quantum gravity,
with an emphasis on the idea that time emerges as a clock-time described by
the matter degrees of freedom.
Sec.~\ref{SEC3} is a novel application of that old idea,
where we argue that such an emergent time in quantum gravity 
provides a natural resolution of the black-hole information paradox.
Finally, the conclusions are drawn in Sec.~\ref{SEC4}.

\section{Time in quantum gravity}
\label{SEC2}

Canonical quantum gravity is based on the Hamiltonian constraint
\begin{equation}\label{eq1}
 {\cal H}\Psi[g,\phi] =0 ,
\end{equation}
where $\cal{H}$ is the Hamiltonian-density operator and $\Psi[g,\phi]$ is the
wave function of the universe, depending on gravitational and matter degrees
of freedom denoted by $g$ and $\phi$, respectively. 
(On the technical level,
the most promissing variant of (\ref{eq1}) is based on loop quantum gravity
\cite{rov}, where $g$ denotes the loop variables.) 
Clearly, $\Psi[g,\phi]$ does not depend on an external time parameter,
which is often referred to as problem of time in quantum gravity
(see, e.g., \cite{isham,kuchar} for older reviews and \cite{rov}
for a review written from a more modern perspective).
Obviously, since $\Psi[g,\phi]$ does not depend on time, the information
encoded in $\Psi[g,\phi]$ cannot depend on time either, i.e., 
information cannot be ``lost''. 
The lack of time dependence can be thought of as ``time evolution''
described by a trivial unitary operator
\begin{equation}\label{eq2}
 U(t)\equiv 1 ,
\end{equation}
which means that the theory is unitary in a trivial sense.
The quantity
\begin{equation}
 \rho[g,\phi]=\Psi^*[g,\phi] \Psi[g,\phi]  
\end{equation}
can be interpreted as probability of given values $g$ and $\phi$, provided
that $\Psi[g,\phi]$ is normalized such that
\begin{equation}
 \int {\cal D}g \, {\cal D}\phi \, \Psi^*[g,\phi] \Psi[g,\phi] =1.
\end{equation}
In loop quantum gravity, the formal measure ${\cal D}g$ 
is mathematically well defined, and there are justified expectations
that ${\cal D}\phi$ could be well defined too.

Even though there is no fundamental notion of time, a phenomenological
notion of time can still be introduced. The most physical way to do it is to
introduce a clock-time \cite{rov}. Essentially, this means that some of the
matter degrees of freedom describe the reading of a ``clock''. In this case the
Hamiltonian $H=\int d^3x \, {\cal H}$ can be split as
\begin{equation}
H=\tilde{H} +H_{\rm clock} ,
\end{equation}
where $H_{\rm clock}$ describes the clock and $\tilde{H}$ is the rest of the Hamiltonian.
The Hamiltonian for a good clock can be approximated by a Hamiltonian of the form
\begin{equation}\label{H_clock}
 H_{\rm clock} \simeq \lambda P_{\rm clock} ,
\end{equation}
where $\phi\equiv \{\tilde{\phi},Q_{\rm clock}\}$, 
$Q_{\rm clock}$ is the configuration variable representing the reading
of the clock, $P_{\rm clock}$ is the canonical momentum conjugated to $Q_{\rm clock}$,
and $\lambda$ is a coupling constant.
Indeed, the resulting classical equation of motion
\begin{equation}
 \frac{dQ_{\rm clock}}{dt}= \frac{\partial H_{\rm clock}}{\partial P_{\rm clock}} \simeq \lambda
\end{equation}
implies
\begin{equation}
 Q_{\rm clock}(t) \simeq \lambda t ,
\end{equation}
so $Q_{\rm clock}$ increases approximately linearly with time, which means that the value of 
$Q_{\rm clock}$ is a good measure of time. 

In quantum theory the momentum
$P_{\rm clock}$ is the derivative operator
\begin{equation}
 P_{\rm clock} = -i\hbar\frac{\partial}{\partial Q_{\rm clock}},
\end{equation}
so (\ref{H_clock}) can be written as
\begin{equation}
 H_{\rm clock} \simeq -i\hbar\frac{\partial}{\partial q_{\rm clock}} ,
\end{equation}
where $q_{\rm clock}\equiv \lambda^{-1} Q_{\rm clock}$.
In this way, (\ref{eq1}) implies a Schrodinger-like equation
\begin{equation}\label{sch}
 \tilde{H}\Psi[g,\tilde{\phi},q_{\rm clock}] \simeq i\hbar\frac{\partial}{\partial q_{\rm clock}} 
\Psi[g,\tilde{\phi},q_{\rm clock}] .
\end{equation}

Even though (\ref{sch}) has the same form as the usual Schr\"odinger equation,
we stress two important differences with respect to the
usual interpretation of time in the Schr\"odinger equation. 
First, $q_{\rm clock}$ is a {\em quantum} observable,
not a classical external parameter. Second, in most cases $q_{\rm clock}$
is a {\em local} quantity, not a quantity that can be associated
with a whole spacelike hypersurface. As we shall see, these two features 
are essential for our resolution of the black-hole information paradox.


\section{Implications on black-hole information paradox}
\label{SEC3}

Now assume that $\Psi[g,\phi]$ is a solution of (\ref{eq1}) that describes
an evaporating black hole. Of course, an explicit construction of such a
solution is prohibitively difficult. Yet, under reasonable assumptions justified
by understanding of semiclassical black holes,
some qualitative features of such a hypothetic solution
can easily be guessed without an explicit solution at hand. In particular, 
it is reasonable to assume
that, at least approximately, the degrees of freedom can be split into inside and outside
degrees of freedom.
Therefore we write
\begin{equation}\label{wf}
\Psi[g,\phi] = \Psi[g_{\rm in},\phi_{\rm in},g_{\rm out},\phi_{\rm out}] .
\end{equation}
This state can also be represented by a pure-state density matrix
\begin{equation}\label{pure}
 \rho[g_{\rm in},\phi_{\rm in},g_{\rm out},\phi_{\rm out}|
g'_{\rm in},\phi'_{\rm in},g'_{\rm out},\phi'_{\rm out}] =
\Psi[g_{\rm in},\phi_{\rm in},g_{\rm out},\phi_{\rm out}]
\Psi^*[g'_{\rm in},\phi'_{\rm in},g'_{\rm out},\phi'_{\rm out}] .
\end{equation}
By tracing out over the inside degrees of freedom,
we get the mixed-state density matrix
\begin{equation}
 \rho_{\rm out}[g_{\rm out},\phi_{\rm out}|g'_{\rm out},\phi'_{\rm out}] =
\int {\cal D}g_{\rm in} \, {\cal D}\phi_{\rm in} \,
\rho[g_{\rm in},\phi_{\rm in},g_{\rm out},\phi_{\rm out}|
g_{\rm in},\phi_{\rm in},g'_{\rm out},\phi'_{\rm out}] ,
\end{equation}
which describes information available to an outside observer.
Next we identify a clock-time of an outside observer, so that
we can write
\begin{equation}
\rho_{\rm out}[g_{\rm out},\phi_{\rm out}|g'_{\rm out},\phi'_{\rm out}] =
\rho_{\rm out}[g_{\rm out},\tilde{\phi}_{\rm out},q_{\rm clock \; out}|
g'_{\rm out},\tilde{\phi}'_{\rm out},q'_{\rm clock \; out}] .
\end{equation}
Finally, by considering the clock-diagonal matrix elements 
$q_{\rm clock \; out}=q'_{\rm clock \; out}\equiv t$, we get an ``evolving''
outside density matrix
\begin{equation}\label{t-evol}
\rho_{\rm out}[g_{\rm out},\tilde{\phi}_{\rm out}|
g'_{\rm out},\tilde{\phi}'_{\rm out}](t) \equiv
\rho_{\rm out}[g_{\rm out},\tilde{\phi}_{\rm out},t|
g'_{\rm out},\tilde{\phi}'_{\rm out},t] . 
\end{equation}
Clearly, the $t$-evolution described by (\ref{t-evol}) may not be unitary.
At times $t$ for which the black hole has evaporated completely, 
(\ref{t-evol}) may correspond to a mixed state, in accordance with 
predictions of the semiclassical theory \cite{hawk2}.
One could think that it is merely a restatement of the information paradox,
but it is actually much more than that. Unlike the standard statement of the paradox
\cite{hawk2}, such a restatement contains also a resolution of the paradox.
Namely,
from the construction of (\ref{t-evol}) it is evident that there is nothing
fundamental about such a violation of unitarity. No information is really lost.
The full information content is encoded in the 
pure state (\ref{pure}) equivalent to the wave function (\ref{wf}).
This is very different from the information loss in the standard formulation
\cite{hawk2}, where information seems to be really lost and no description
in terms of pure states seems possible.

To see more explicitly where the information is hidden, it is useful
to introduce {\em two} clocks, such that (\ref{wf}) can be written as
\begin{equation}\label{wf2}
\Psi[g_{\rm in},\phi_{\rm in},g_{\rm out},\phi_{\rm out}] =
\Psi[g_{\rm in},\tilde{\phi}_{\rm in},q_{\rm clock \; in},
g_{\rm out},\tilde{\phi}_{\rm out},q_{\rm clock \; out}].
\end{equation}
Here $q_{\rm clock \; in}$ and $q_{\rm clock \; out}$ are configuration
variables describing an inside clock and an outside clock, respectively.
Assuming that the black hole eventually evaporates completely,
the inside clock cannot show a time larger than some value $t_{\rm evap}$
corresponding to the time needed for the complete evaporation.
More precisely, the probability that $q_{\rm clock \; in}>t_{\rm evap}$
is vanishing, so 
\begin{equation}\label{wf3}
 \Psi[g_{\rm in},\tilde{\phi}_{\rm in},q_{\rm clock \; in},
g_{\rm out},\tilde{\phi}_{\rm out},q_{\rm clock \; out}]  =0 \;\;\;
{\rm for} \;\;\; q_{\rm clock \; in}>t_{\rm evap} .
\end{equation}
The existence of the wave function (\ref{wf2}) implies that 
the system can be described by a pure state even after the complete evaporation.
However, this description is trivial, because (\ref{wf3}) says that the wave function
has a vanishing value
for $q_{\rm clock \; in}>t_{\rm evap}$. Still, even a nontrivial pure-state description for 
$q_{\rm clock \; out}>t_{\rm evap}$ is possible, provided that $q_{\rm clock \; in}$ 
is restricted to the
region $q_{\rm clock \; in}<t_{\rm evap}$. In this case (\ref{wf2}) describes the
correlations between the outside degrees of freedom after the complete evaporation and 
the inside degrees of freedom before the complete evaporation. 
In other words, if one asks where the 
information after the complete evaporation is hidden, then the answer is -- 
{\em it is hidden in the past}.
Of course, experimentalists cannot travel to the past, so information is lost for the experimentalists.
Yet, this information loss is described by a pure state, so one does not need to use the
Hawking formalism \cite{hawk2} in which a state evolves from a pure to a mixed state.
By avoiding this formalism one avoids its pathologies \cite{banks} too, which may be 
viewed as the main advantage of our approach. 

One might object that information hidden in the past is the same as 
information destruction, but it is not.
The difference is subtle and essential for our approach, so let us explain 
it once again more carefully. Information hidden in the past and information destruction
are the same for an observer who views the world as an entity
that evolves with time $t$ in (\ref{t-evol}). However, such a view of the world
is emergent rather than fundamental, because time is emergent
rather than fundamental. At the fundamental level there is no time
and no evolution. The fundamental world is static and unitary, as described by
(\ref{eq2}). 
The concept of ``past'' refers to something which does not longer exist at the
emergent level, but it still exists at the fundamental level.
Thus, at the fundamental level, information
is better described as being {\em present} in the past and only {\em hidden}
for an emergent observer, rather than being destroyed.
In this sense, our resolution of the information paradox does not remove
the non-unitary time evolution entirely. Instead,  
{\em it shifts the non-unitary time evolution from a fundamental level to an emergent one}.

One might still argue that we have only shifted the problem (from one level
to another) and not really solved it. But in our view such a shift of the problem
is also a solution, or at least a crucial part of a solution.
Namely, it is typical for emergent theories in physics that they lack full self-consistency,
even when the underlying fundamental theories are self-consistent.
Indeed, a presence of an inconsistency in an otherwise successfull physical theory 
is often a sign that this theory is not fundamental, but 
emergent. (A classic example is the ultraviolet catastrophe in classical statistical
mechanics. It was resolved by Planck and others by recognizing that 
classical statistical mechanics emerges from 
more fundamental quantum statistical mechanics, which does not involve
the ultraviolet catastrophe. In this way the inconsistency of classical statistical
mechanics was not removed, but shifted from a fundamental to an emergent
level.)
In our case of the black-hole information paradox, the emergent theory is not self-consistent 
as it violates unitarity. We resolve the problem by identifying a more fundamental unitary theory
from which the unitarity-violating theory emerges. 
The unitarity violation is nothing but a sign 
that {\em the emergent description in terms of time evolution is not fully applicable
to the phenomenon of black-hole evaporation,
and the fundamental theory involving no time evolution is a more appropriate
description.}
In our opinion, it is a legitimate resolution
of the black-hole information paradox, even if an unexpected one.

\section{Conclusion}
\label{SEC4}

To conclude, canonical quantum gravity lacks a fundamental notion
of time evolution, which implies trivial unitarity of the theory at the fundamental level.
Time and evolution are emergent concepts, defined with the aid of a physical clock.
In general, such a clock-time only has a local meaning and is represented 
by a quantum observable. Therefore, information present on a global
spacelike hypersurface does not play any fundamental role. Consequently,
even if observers living after the complete evaporation of a black hole
cannot see all information encoded in the wave function of the universe,
which can be interpreted as effective violation of unitarity for the observers, 
the full wave function of the universe still contains all the information
and no fundamental violation of unitarity takes place.
In this way both fundamental unitarity and phenomenological 
information loss may peacefully coexist,
which resolves the black-hole information paradox.

\section*{Acknowledgements}

%
%
This work was supported by the Ministry of Science of the
Republic of Croatia under Contract No.~098-0982930-2864.

\end{document}